\documentclass[pra,aps,showpacs,groupedaddress,nofootinbib,twocolumn]{revtex4-1}

\usepackage{amsmath,amsfonts,amssymb,amsthm}
\usepackage{wrapfig}
\usepackage{graphicx}
\usepackage{bbm}
\usepackage{graphics}
\usepackage{color}

\def \beq {\begin{equation}}
\def \eeq {\end{equation}}
\def \ba {\begin{eqnarray}}
\def \ea {\end{eqnarray}}
\def \a {\hat{a}}
\def \ad {\hat{a}^{\dag}}
\def\ket#1{\left| #1\right>}
\def\bra#1{\left< #1\right|}

\newcommand{\vek}[1]{\mathbf #1}
\newcommand{\anni}{\hat a}
\newcommand{\crea}{{\hat a}^{\dagger}}
\newcommand{\abs}[1]{\left\vert#1\right\vert}

\usepackage{ulem}
\begin{document}




\title{Fast and robust quantum computation with ionic Wigner crystals}




\author{J. D. Baltrusch$^{1,2,3}$, A. Negretti$^1$, J. M. Taylor$^4$ and T. Calarco$^{1,5}$}
\affiliation{$^1$Institute for Quantum Information Processing, University of
  Ulm, Albert-Einstein-Allee 11, D-89069 Ulm, Germany\\
$^2$Grup d'\`Optica, Edifici CC, Universitat Aut\`onoma de Barcelona (UAB), 
E-08193 Bellaterra (Barcelona), Spain \\
$^3$Theoretische Physik, Universit\"at des Saarlandes, 66041
Saarbr\"ucken, Germany\\
$^4$Joint Quantum Institute and the National Institute of Standards and Technology, College Park, Maryland 20742, USA\\
$^5$Department of Physics, Harvard University, and ITAMP, Cambridge,
MA  02138, USA}
\date{\today}




\begin{abstract}
We present a detailed analysis of the modulated-carrier quantum phase gate implemented 
with Wigner crystals of ions confined in Penning traps. We elaborate on a recent scheme, 
proposed by two of the authors, to engineer two-body interactions between ions in such crystals. 
We analyze for the first time the situation in which the cyclotron ($\omega_{\mathrm{c}}$) 
and the crystal rotation ($\omega_{\mathrm{r}}$) frequencies do not fulfill the condition 
$\omega_{\mathrm{c}} = 2 \omega_{\mathrm{r}}$. It is shown that even in the presence of 
the magnetic field in the rotating frame the many-body (classical) Hamiltonian describing small 
oscillations from the ion equilibrium positions can be recast in canonical form. As a 
consequence, we are able to demonstrate that fast and robust two-qubit gates are 
achievable within the current experimental limitations. Moreover, we describe a realization 
of the state-dependent sign-changing dipole forces needed to realize the investigated 
quantum computing scheme. 
\end{abstract}





 
\maketitle




\section{Introduction}
\label{sec:intro}

Despite the huge experimental progress to cool, trap, and manipulate 
single particles such as atoms and molecules at the quantum level, the 
way to build up a quantum computing hardware working with several 
hundreds of quantum bits (qubits) in a coherent and controllable 
manner is still long. By means of quantum optimal control techniques it is possible, 
at least theoretically, to perform one- and two-qubit quantum gates with fidelities  
above the demanding thresholds of fault-tolerant quantum computation 
\cite{Charron2002,Palao2002,Calarco2004,Treutlein2006,Charron2006,
Montangero2007,Motzoi2009,Safaei2009,Poulsen2010}. 
These thresholds fix an error between 0.01\% to fractions of a percent 
\cite{Steane2003,Knill2005}. Up to now, only with cold trapped ions quantum 
gates with a fidelity of 99.3\% have been experimentally demonstrated 
\cite{Leibfried2003,Benhelm2008}, which is not too far from the aforementioned 
thresholds. Similar fidelities have been also obtained for small quantum algorithms 
\cite{Gulde2003,Chiaverini2005}. 

Nowadays, however, most of the experimental 
efforts of the atomic and molecular physics community are concentrated in the 
design and fabrication of microtraps, both for ions \cite{Kielpinski2002,Stick2006} 
and neutral atoms \cite{Folman2000,Reichel1999}. 
Even though these efforts are important, significant technical issues related
to the miniaturization and trapping methodologies arise when
scaling to many particles, and therefore new strategies have to be devised. 
A possible solution to the problem is the separation between the qubits used
as quantum memory and the ones employed to process the information \cite{Oskin2002} 
or, alternatively, the exploitation of quantum distributed networks \cite{Cirac1997}. 
Another approach, instead, consists in the use of collective states of atomic ensembles with a 
multilevel internal structure as qubits \cite{Brion2007}.

Apart from these technological efforts and alternative solutions, nobody can yet 
say which of the various physical implementations will be the successful one. 
It is fair to say, however, that ions represent a good candidate to implement 
a multi-qubit quantum processor. Indeed, two-qubit gates with ions can be realized in 
about few tens of $\mu$s \cite{Steane2000,Chen2006}, and qubits 
stored in internal electronic degrees of freedom of an ion have coherence lifetimes 
ranging from 1 s to 100 s or more~\cite{Chen2006}.

\begin{figure}[t]
\begin{center}
\includegraphics[width=2.4in]{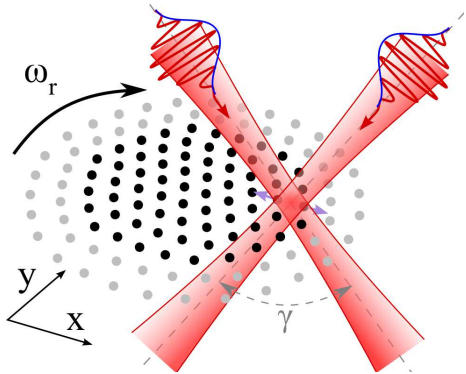}
\end{center}
\caption{(Color online). Two-dimensional Coulomb crystals of ions in a Penning trap 
rotating at frequency $\omega_{\mathrm r}$. To manipulate the internal states of the 
ions, laser beams can address single sites or multiple ions.}
\label{fig:artistsview}
\end{figure}

Coulomb --- also named classical Wigner --- crystals confined in Penning 
traps are natural candidates for a quantum memory, since the separation among 
ions, about  10 $\mu$m, allows to individually manipulate their internal degrees of 
freedom. Such a trap scheme uses static electric fields to confine charge particles 
in the axial direction (the $z$ axis in Fig. \ref{fig:artistsview}), whereas the radial 
confinement is provided by a strong uniform magnetic field along the axial 
direction. Currently, Penning traps allow to trap up to $10^8$ ions \cite{Anderegg2010}. An appropriate 
choice of the trap parameters (e.g., tight axial confinement) allows the ionic ensemble 
to crystallize in a two-dimensional (2D) hexagonal lattice configuration with an inter-particle 
spacing on the order of tens of $\mu$m \cite{Mitchell1998}, and therefore to manipulate a large 
number of qubits without specific micro-trap designs. The high phonon mode density, 
however, does not permit to resolve single modes for sideband cooling. Hence, 
Doppler and sympathetic cooling are the most natural techniques to be employed; 
we also note that Sisyphus cooling might be an alternative methodology \cite{Wineland1992}.
Current experiments, however, performed with Doppler cooling, can reach temperatures of few mK 
\cite{Mitchell2001}, that is, a high thermal occupation number distribution of phonon 
modes. Nonetheless, efficient quantum computation and production of small 
cluster states are theoretically possible \cite{Porras2006,Taylor2008}, and recently 
full control of the qubit Bloch vector with $\sim$99.85 \% fidelity for Rabi flopping 
has been experimentally demonstrated~\cite{Biercuk2009}.

The two-qubit gate scheme considered in the proposals of Refs.
\cite{Porras2006,Taylor2008} is based on the so called 
``pushing gate'' (or its variant, the modulated-carrier gate), 
where a spatially inhomogeneous laser field together with an appropriate 
combination of polarizations and frequencies induces a state-dependent 
dipole force on two nearest neighbours of the 2D Coulomb crystal (see 
Fig. \ref{fig:artistsview}). Depending on the configuration of 
lasers and polarizations the displacements of the ions away from their
equilibrium positions can be either perpendicular to the plane 
of the crystal \cite{Porras2006} or along the in-plane separation of the ions 
\cite{Cirac2000,Calarco2001,Sasura2003,Taylor2008}. 
The coupling between these displacements, mediated by phonons, yields 
entanglement of the internal states (qubits) of the ions, that is, the 
desired quantum gate between ions. 

In addition to the confinement, the radial electric and axial magnetic 
fields induce a drift that causes in-plane rotation of the crystal (see Fig. \ref{fig:artistsview}), whose 
frequency $\omega_{\mathrm{r}}/(2\pi)$ is typically on the order of few tens 
of kHz \cite{Mitchell1998}. There are two possible solutions to our quantum hardware design: 
either we use a co-rotating (with the crystal) laser beam in order to 
realize the desired two-qubit quantum gate, or we have to perform the gate in a 
time, $\tau_{\mathrm{g}}$, such that the crystal rotation has a negligible effect 
on the gate operation. The latter solution translates in the condition 
$\omega_{\rm r}\tau_{\mathrm{g}}/(2\pi) \ll 1$. Such a requirement is 
instrumental because, in order to accumulate the necessary two-ion 
phase for the quantum gate we aim to implement, the ions have to 
experience the applied light force for the entire gate operation, or else, 
the required phase would be achieved only partially. 

While the former solution applies for all rotation frequencies, but relies on a 
more sophisticated experimental setup, the latter restricts the range of possible 
values of $\omega_{\mathrm{r}}$. Thus, in both proposals  \cite{Porras2006,Taylor2008}, where 
the rotation and cyclotron frequencies fulfill $2\,\omega_{\rm r} = \omega_{\rm c}$, 
the aforementioned condition is satisfied when $\tau_{\mathrm{g}}$ is on the 
order of ns, whereas the modulated-carrier gate of Ref. \cite{Taylor2008} 
had  $\tau_{\mathrm{g}}=$ 5 $\mu$s. Given the above, such a proposal requires a 
co-rotating laser beam. Thus, by maintaining $2\,\omega_{\rm r} = \omega_{\rm c}$ 
one should reduce $\omega_{\rm c}$. This approach, however, would not help since 
the smaller the cyclotron frequency is, the longer the gate operation. 
Instead, if we abandon the assumption $2\,\omega_{\rm r} = \omega_{\rm c}$ and look 
at moderate rotation frequencies, at the expenses of possible large modulations of 
the force, we are able to fulfill $\omega_{\rm r}\tau_{\mathrm{g}}/(2\pi) \ll 1$. 
Additionally, low rotating frequencies result in low densities and large inter-particle 
spacing, and therefore in an easier way to address the trapped ions  with a laser field.

Thus, the main goal of this work is to analyze this regime and, at the same time, to perform 
robust two-qubit gates within a range of experimentally achievable temperatures. 

In the following we shall we present the general theory of the modulated-carrier 
push two-qubit gate (Sec. \ref{sec:modcargate}) with details that 
were briefly mentioned in Ref. \cite{Taylor2008}. Subsequently, in Sec. \ref{sec:coupling}, 
we investigate the situation in the presence of the magnetic field in the rotating frame 
of reference and the relative gate performance. In section \ref{sec:force} we describe how to 
physically realize the state-dependent force required for the proposed quantum 
processor, and Sec. \ref{sec:conc} summarizes our results and provides some 
future prospectives. 
 
\section{Modulated-carrier gate}
\label{sec:modcargate}

In the following we make the approximation that the 
Wigner crystal is a rigid body, which is a good approximation in the
magnetohydrodynamic regime (one component plasma) or at 
equilibrium \cite{Dubin1999}. Hence, in a rotating frame, 
the Hamiltonian of a crystal with $N$ ions written in cylindrical 
coordinates [$\vec{r}\equiv(r,\theta,z)$] is given by \cite{Dubin1999}

\begin{eqnarray}
H_R(\omega) &=& \sum_{k=1}^N \left\{\frac{p^2_{r_k} + p^2_{z_k}}{2 m} 
+ \frac{[p_{\theta_k} - m(\omega_{\mathrm c} - 2 \omega)r_k^2/2]^2}{2 m r_k^2}
\right\} 
\nonumber\\
&+& \sum_{k=1}^N \left\{\Upsilon(2
z_k^2 - r_k^2) + \frac{m}{2}\omega(\omega_{\mathrm c} - \omega)r_k^2\right\} 
+ V_{\mathrm{c}},\nonumber\\
\label{eq:Homega}
\end{eqnarray}
with $m$ being the mass of the ion and

\begin{eqnarray}
V_{\mathrm{c}} = \frac{e^2}{4 \pi \epsilon_0} \sum_{k<j}\frac{1}{\vert
  \vec{r}_k - \vec{r}_j\vert}.
\end{eqnarray}
Here $\Upsilon$ is a parameter describing the trap geometry and applied voltage 
on the electrodes \cite{Gosh1995}, $\epsilon_0$ the vacuum permittivity, $e$ the 
electron charge, and $\omega_{\mathrm c} = e B / m$ is the cyclotron frequency.
We see, from the first line of Eq.~(\ref{eq:Homega}), that there exists a special rotating 
frame, $\omega = \omega_{\mathrm c}/ 2$, such that the minimal coupling disappears, 
and, in this section, we shall consider such a frame of reference together with 
$\omega_{\mathrm r} = \omega_{\mathrm c}/ 2$ (i.e., the frame of reference coincides 
with the crystal). We note, that with "minimal coupling" we refer to the interaction 
$\vec{p} \cdot\vec{A}$. Such terminology is typically used in quantum field theory 
\cite{Itzykson1980}.  

Finally, it is worth to remind that the gate we aim to accomplish realizes the true table 
$\ket{\epsilon_1,\epsilon_2}\rightarrow e^{i\theta\epsilon_1\epsilon_2}
\ket{\epsilon_1,\epsilon_2}$ with $\epsilon_{1,2} = 0,1$ and 
$\theta = \theta_{00} - \theta_{01} - \theta_{10} + \theta_{11}$ \cite{Calarco2001,Sasura2003}. 
Specifically, we are interested in a phase gate with $\theta=\pi$, which, up to 
additional single-qubit rotations, is tantamount to a two-qubit controlled NOT 
gate~\cite{Chen2006}. 

\subsection{Normal modes and canonical quantization}
\label{sec:phonons}

The Hamiltonian (\ref{eq:Homega}) in cartesian coordinates [$\vec{r}\equiv(x,y,z)$] reduces to

\begin{eqnarray}
H_R\left(\frac{\omega_{\mathrm c}}{2}\right) \!=\! \sum_{k=1}^N \left\{\frac{\vec{p}^{\,2}_{k}}{2 m} 
+ \frac{m}{2}\left[ \omega_z^2 z_{k}^2 + \omega_{xy}^2(x_{k}^2 +
  y_{k}^2)\right] \right\} + V_{\mathrm{c}}, \nonumber\\
\label{eq:Homega-special}
\end{eqnarray}
where $\omega_z = \sqrt{4\Upsilon / m}$ is the axial frequency, and 
$\omega_{xy} = 1/2 (\omega_{\mathrm c}^2 - 2\omega_z^2)^{1/2}$ the in-plane one. 

By performing a Taylor expansion of the potential up to second order around the stable
equilibrium configuration, obtained by minimizing the total crystal energy, we can express 
the Hamiltonian in the new coordinates $q_{n,\eta} \equiv \eta_{n} - \eta_{n}^0$, that is, the 
displacements from the equilibrium positions. Hence, it is possible to determine 
an orthogonal transformation $M$ such that\footnote{Hereafter we shall use latin symbols for index
  the ions and greek symbols for the cartesian coordinate of the force vector
  acting on the ions.}
  
\begin{equation}
H_R\left(\frac{\omega_{\mathrm c}}{2}\right) \approx \sum_{n,\eta} \left\{\frac{P_{n,\eta}^2}{2m} 
+ \frac{m}{2} \omega_{n,\eta}^2 Q_{n,\eta}^2\right\}
\end{equation}
with $Q_{n,\eta} = \sum_{k,\mu} M_{n,\eta;k,\mu} q_{k,\mu} [ = M \bf{q}]$,
and $P \equiv p$. 

Now, we perform the canonical quantization and we introduce the 
creation (annihilation) operators $\crea_K$ ($\anni_K$) for each mode 
$K\equiv (n,\eta)$, along with the harmonic oscillator ground state size 
$\alpha_K = \sqrt{\hbar/m \omega_K}$. Hence, the (phononic) Hamiltonian 
operator reads

\begin{equation}
\hat H_R = \sum_{K} \hbar \omega_{K} (\crea_{K} \anni_{K}^{\phantom{\dagger}} + 1/2),
\end{equation}
where for the sake of simplicity we drop $\left(\frac{\omega_{\mathrm c}}{2}\right)$ 
in $\hat H_R$.

\subsection{Adiabatic and oscillatory quantum gates}
\label{sec:adiafastgate}

Let us consider a spatially inhomogeneous laser field appropriately detuned 
from the internal states such that it produces a state-dependent displacement of the 
ions. Then, the matter-field interaction, in the electric dipole approximation, becomes 

\begin{eqnarray}
  \!\hat V & = & \sum_{j=1}^N [\hat{\vec{q}}_j \cdot \vec{f}_j(t)] \hat{\sigma}^z_j 
= \sum_K \frac{\alpha_K}{\sqrt{2}} \hat{f}_K(t) (\ad_K + \a_K),
\end{eqnarray}
where $\vec{f}_j$ is the three dimensional force due to the gradient in the laser
intensity, and $\hat{\sigma}^z_j$ is the $z$ Pauli matrix. Here 
the following relation for the displacement coordinate operator 

\begin{equation}
\hat{q}_K = \sum_{K^{\prime}} M_{K^{\prime}; K} \hat{Q}_{K^{\prime}} 
=\sum_{K^{\prime}} M_{K^{\prime}; K} \frac{\alpha_K}{\sqrt{2}} (\a_K +
\ad_K)
\label{eq:normalmode}
\end{equation}
has been used. Thus, we have [$K\equiv (j,\mu)$]

\begin{eqnarray}
 \hat{f}_K(t) = \sum_{j^{\prime},\mu^{\prime}} M_{K;j^{\prime},\mu^{\prime}}[\vec{f}_{j^{\prime}}(t)]_{\mu^{\prime}}\hat{\sigma}^z_{j^{\prime}},
\end{eqnarray}
where $[\vec{f}_{j^{\prime}}(t)]_{\mu^{\prime}}$ is the $\mu^{\prime}=x,y,z$ component of the 
three-dimensional vector $\vec{f}_{j^{\prime}}(t)$. 
Hence, the full problem reduces to $3 N$ independent, driven oscillators.

When the temporal profile of the force fulfills the condition ${\rm lim}_{t \rightarrow \pm \infty} f(t) = 0$, 
the unitary time evolution operator is given by 
$\hat U_K(t) = e^{-i \phi_K(t)} \exp(\beta_K \ad_K - \beta_K^* \a_K)\exp(-i\omega_Kt\,\ad_K\a_K)$, where $\phi_K$ 
and $\beta_K$ satisfy the differential equations~\cite{Garcia2003,Garcia2005}

\begin{equation}
\dot{\beta}_K = -i \omega_K \beta_K + i \frac{\alpha_K}{\hbar
\sqrt{2}} \hat{f}_K(t), \; \dot{\phi}_K = \frac{\alpha_K}{\hbar \sqrt{2}}
\hat{f}_K(t) {\mathrm{Re}}[\beta_K(t)].
\label{eq:difeq-betaphi}
\end{equation}
Given that, let us consider the adiabatic regime regime where $\hat{f}_K(t)$ varies 
slowly with respect to $\omega_K$ \cite{Calarco2001}. Adiabatic elimination, by taking 
$\dot{\beta}_K \rightarrow 0$, yields

\begin{eqnarray}
  \beta_K \approx \frac{\alpha_K \hat{f}_K(t) }{\hbar \omega_K \sqrt{2}}, \qquad
  \dot\phi_K \approx \frac{\alpha_K^2 \hat{f}_K^2(t)}{2 \hbar^2 \omega_K}.
\end{eqnarray}
Thus, the displacement of a normal mode $K$ induced by the gate is
proportional to the force applied, and can be made zero independent of
the initial phonon state by starting and ending with zero force. This
eliminates any potential error due to entanglement between phonons and
the internal states of the ions.  Similarly, the overall phase
accumulated $\sum_K \phi_K(\tau_{\mathrm{g}})$ does not depend on the initial phonon
state. However, for a gate occurring over a time interval $[0,\tau_{\mathrm{g}}]$, the
final qubit state has applied $\exp(-i \sum_{nj}\phi_{nj}\hat\sigma^z_n\hat\sigma^z_j)$, 
where the two-particle phases arise from 

\begin{equation}
  \hat f_K^2(t) = \sum_{j,n;\mu,\eta} M_{K;j,\mu} M_{K;n,\eta} 
  [\vec{f}_j(t)]_{\mu}[\vec{f}_n(t)]_{\eta} \hat\sigma^z_j \hat\sigma^z_n.
\end{equation}
Thus, the two-particle phase is given by

\begin{equation}
  \phi_{nj} = \sum_{\mu,\eta} S^{(nj)}_{\mu\eta}
  \int_0^{\tau_{\mathrm{g}}}\! dt\, [\vec{f}_j(t)]_{\mu}[\vec{f}_n(t)]_{\eta},
\label{eq:adiabatic}
\end{equation}
where the term outside the integral is a shape independent form factor, whose 
specific form is given by

\begin{equation}
S^{(nj)}_{\mu\eta}  = \sum_K \frac{\alpha_{K}^2}{2 \hbar^2\omega_{K}} M_{K;j,\mu} M_{K;n,\eta}.
\label{eq:S-adiabatic}
\end{equation}
Hence, we can think about (\ref{eq:adiabatic}) as a convolution of the 
forces on the two particles, modified by the form factor representative 
of the characteristic oscillator variance over its frequency, which is 
overall proportional to $\omega_K^{-1}$. 

Now, let us consider a scheme with a force $f(t) \rightarrow \cos(\nu t) f(t)$, 
where the carrier frequency $\nu$ must be much larger than the modes 
of frequency $\omega_K$ that are coupled to the force (this averages out any net 
displacement). If the modulation 
$f(t)$ is slow as compared to $\nu$ (but with no restriction with respect 
to $\omega_K$), we can perform a similar adiabatic elimination as above, 
and get a gate with the same desirable properties that can operate 
non-trivially on arbitrarily in-plane vibrational modes at very high 
temperatures.

For adiabatic elimination with respect to $\nu$, we choose the Ansatz 
$\beta_K = \beta_K^+ e^{i \nu t} + \beta_K^- e^{-i \nu t}$ for each mode. By
inserting this Ansatz into the differential equation (\ref{eq:difeq-betaphi}) 
we obtain

\begin{eqnarray}
  \dot{\beta}_K^+ & = & e^{-2 i \nu t}\left[ i \frac{\alpha_K}{2 \sqrt{2}\hbar}
    \hat f_K(t)  - \dot{\beta}_K^- - i(\omega_K - \nu) \beta_K^-\right]
   \nonumber \\
& & \ \ + i \frac{\alpha_K}{2\sqrt{2}\hbar} \hat f_K(t) -i(\omega_K + \nu) \beta_K^+.  
\end{eqnarray}
Separate adiabatic elimination of $\beta_K^-$ and $\beta_K^+$
yields $\beta_K^{\pm} = \alpha_K \hat f_K(t)/[2 \sqrt{2} \hbar(\omega_K \pm
\nu)]$. As before, in the pure adiabatic regime, we find that 
the displacement of a normal mode induced by the gate is proportional 
to the force applied. Again, it can be made zero independent of 
the initial phonon state by starting and ending with zero
force, and therefore eliminating any potential error due to entanglement
between phonons and the internal states of the ions.

Now, we examine the two-particle phase induced in this new
scenario. The time evolution of the phase is governed by 
\cite{Taylor2008}

\begin{equation}
  \dot{\phi}_K = \frac{\alpha_K^2}{2 \hbar^2}
  \frac{\omega_K}{(\omega_K^2-\nu^2)}
  \cos^2(\nu t)  \hat f_K^2(t),
\end{equation}
where the quickly varying component $\cos^2(\nu t)$ can be replaced with
$1/2$. As described in the adiabatic regime, the overall
phase accumulated $\sum_K \phi_K(\tau_{\mathrm{g}})$, for a gate occurring over a
time interval $[0,\tau_{\mathrm{g}}]$, does not depend on the phonon initial state. 
In this case the pulse-shape independent form factor is given by 
\cite{Taylor2008}

\begin{equation}
S^{(nj)}_{\mu\eta}  = -\sum_K \frac{\alpha_{K}^2\omega_K}
{4 \hbar^2(\nu^2 - \omega_K^2)} M_{K;j,\mu} M_{K;n,\eta}.
\label{eq:Sfastall}
\end{equation}
Performing a Taylor expansion in $1/\nu^2$ the first term 
is proportional to $\sum_K M_{K;j,\mu} M_{K;n,\eta}= \delta_{j,n}\delta_{\mu,\eta}$ 
($\delta_{j,n}$ indicates the Kronecker symbol). This follows from the fact that 
$M$ is an orthogonal matrix. Physically, this arises due to the coherent 
averaging of in-phase oscillating ions without any virtual excitation 
of phonons---accordingly, no two-body phase should be expected. The 
second term of the expansion is non-zero and yields

\begin{equation}
 \tilde{S}^{(nj)}_{\mu\eta}  = - \frac{1}{4 \hbar m \nu^4} 
\sum_K\omega_K^2 M_{K;j,\mu} M_{K;n,\eta}
+ O\left(\nu^{-6}\right).
\label{Sfast}
\end{equation}
Compared to adiabatic push gates, the modulated-carrier gate is
inverted in sign and it is multiplied (in phase) by a factor $(\omega_K/\nu)^4/2$ 
[see Eq. (\ref{eq:S-adiabatic})]. In the case of a lateral operating modulated-carrier 
gate with $\omega_{xy}\ll\nu\ll\omega_z$, the accumulated phase is enhanced by 
a factor $(\omega_z/\nu)^4/2$ with respect to an adiabatic push gate with a force 
moving the ions in the axial ($z$) direction for the same 
laser parameters. Given that, the gate time needed to perform a $\pi$-phase gate 
is reduced. In the opposite case, that is, for an adiabatic in-plane push gate 
($\omega_K \sim \omega_{xy}$), and for the same laser parameters, the lateral 
modulated-carrier gate is reduced in phase, and therefore a longer $\tau_{\mathrm{g}}$ 
is required. Thus, compared to the proposal of Ref. \cite{Porras2006}, where the push gate 
operates in the axial direction, our modulated-carrier gate working with in-plane modes 
yields a larger two-ion phase for a given set of laser parameters, and therefore it enables to 
perform a larger number of quantum gates within the coherence time of the system.

\section{Modulated-carrier gate with minimal coupling}
\label{sec:coupling}

In this section we analyze the situation where $\omega_{\mathrm r} \ne
\omega_{\mathrm c}/2$, for which we have three reasonable choices for the 
rotating frame of reference: 

\begin{itemize}
\item  $F_1$ coincides with the lab frame, where the equilibrium positions 
  of the ions in the crystal are time-dependent and the minimal coupling does 
  not vanish;
\item  $F_2$ rotates with a frequency $\omega = \omega_{\mathrm c}/2$, as
  in the previous section, where the minimal coupling vanishes, but the 
  equilibrium positions are time-dependent;
\item  $F_3$ rotates with a frequency $\omega = \omega_{\mathrm r}$, where 
  equilibrium positions are time-independent, but the minimal coupling does 
  not vanish.
\end{itemize}

\subsection{Equilibrium configuration of the crystal}
\label{sec:equpo}

Let us discuss which of the frames of reference $F_{1,2,3}$ 
is more suitable to numerically determine the equilibrium 
configuration of the system for a fixed (a priori) value of 
total canonical angular momentum $P_{\theta}
$\footnote{When $\omega_{\mathrm r} \ne \omega_{\mathrm c}/2$, 
the total canonical angular momentum $P_{\theta}\ne 0$, but
it is still a constant of motion \cite{Dubin1999}.}.
Since we are not concerned with relativistic velocities, 
the electromagnetic fields involved in the problem are the same in all 
frames of reference. Consequently, the angular momentum of an ion 
in a frame rotating with uniform angular velocity with respect to the 
(inertial) laboratory frame coincides with the one in the latter 
\cite{Fasano2006}. This conclusion allow us to find the equilibrium 
configuration of the crystal, for a given value of $P_{\theta}$, by 
choosing a frame of reference rotating with angular velocity 
$\omega = \omega_{\mathrm c}/2$ (the frame $F_2$ in the above 
outlined list) in such a way that the coordinate systems at the initial 
time $t=0$ of $F_2$ and $F_3$ do coincide. 
Such a choice simplifies the numerical minimization procedure, 
because the minimal coupling in the (classical) Hamiltonian vanishes. 
We underscore, however, that $F_2$ is utilized only at 
time $t=0$ for the determination of the equilibrium configuration of the
crystal. Instead, for times $t>0$ we use $F_3$, where the
equilibrium positions are time-independent. With such a choice 
the numerical effort in order to assess the gate performance 
is significantly reduced.

Besides this, we also note that not all rotation frequencies $\omega_{\mathrm{r}}$ 
of the crystal allow to have a stable configuration, that is, ions confined within 
a well-defined spatial region. Indeed, by rewriting the addend of the second 
sum in Eq.~(\ref{eq:Homega}) as

\begin{eqnarray}
\Upsilon(2z_k^2 - r_k^2) 
+ \frac{m}{2}\omega_{\mathrm{r}}(\omega_{\mathrm c} - \omega_{\mathrm{r}})r_k^2
= \frac{m \omega_z^2}{2}(z_k^2 + \beta r^2_k)\nonumber\\
\end{eqnarray}
we see that the potential is confining if and only if $\beta$ is positive. Here the 
anisotropy parameter $\beta$ is defined as

\beq
\beta = \frac{\omega_{\mathrm{r}}(\omega_{\mathrm{c}}-\omega_{\mathrm{r}})}{\omega_z^2} 
- \frac{1}{2}
= \frac{\omega_{\mathrm{r}}}{\alpha_z^2\omega_{\mathrm{c}}}\left(
1 - \frac{\omega_{\mathrm{r}}}{\omega_{\mathrm{c}}}\right) - \frac{1}{2},
\label{eq:beta}
\eeq
where $\alpha_z = \omega_z/\omega_{\mathrm{c}}$. Importantly, $\beta$ relies 
only on $\alpha_z$ and the ratio $\omega_{\mathrm{r}}/\omega_{\mathrm{c}}$. 
Thus the range of admissible frequencies is: $\omega_{\mathrm{m}} < 
\omega_{\mathrm{r}} < \omega_{\mathrm{c}} - \omega_{\mathrm{m}}$, where
$\omega_{\mathrm{m}} = \omega_{\mathrm{c}}/2 - \omega_{xy}$ is the
magneton frequency \cite{Dubin1999}. Of course, the admissible regime 
is also constrained by the condition $\alpha_z <1/\sqrt{2}$. In order to access 
lower rotation frequencies, the trap parameters might be changed by increasing 
$\omega_{xy}$, that is, by lowering $\omega_z$. Attention has to be paid, however, 
when $\omega_{\mathrm{c}}$ and $\omega_z$ are changed, since due to such a manipulation 
different structural phase transitions may occur. In particular, 
we are interested in the limit $\beta\ll 1$, where a 2D hexagonal lattice structure
appears~\cite{Dubin1999}. 

\subsection{Quadratic expansion of the Hamiltonian}
\label{sec:quadraticExp}

Let us introduce the typical scale of length $\ell_s$, momentum $p_s$,
and energy $E_s$ in our problem:

\begin{eqnarray}
\ell_s = \left(
\frac{e^2}{4\pi\epsilon_0 m \omega^2_{\mathrm{c}}}
\right)^{\frac{1}{3}} 
\,\,\,\,\, p_s = \ell_s m \omega_{\mathrm{c}} 
\,\,\,\,\, E_s = \frac{e^2}{4\pi\epsilon_0\ell_s}.
\end{eqnarray}
Then, the Hamiltonian (\ref{eq:Homega}) in cartesian coordinates becomes

\begin{eqnarray}
H_R(\omega) &=& \frac{1}{2}\sum_{k=1}^N \left[p^2_{x_k} + p^2_{y_k} + p^2_{z_k}
+(y_k p_{x_k} - x_k p_{y_k})\times\right. 
\nonumber\\
&\times& \left.(1+2\alpha)\right]
+\frac{1}{4}\sum_{k=1}^N\left[\alpha^2_z(2 z_k^2 - r_k^2) + \frac{r^2_k}{2}\right] \nonumber\\
&+& \frac{1}{2}\lim_{\epsilon \rightarrow 0}\sum_{k,j=1}^N\frac{1 - \delta_{k,j}}{\vert\vec{r}_k -
  \vec{r}_j + \epsilon\vert},
\label{eq:Homega-dl}
\end{eqnarray}
where the substitutions $H_R(\omega)\rightarrow H_R(\omega)/E_s$,
$(r, z)\rightarrow (r, z)/\ell_s$, $(p_x, p_y, p_z) \rightarrow (p_x,
p_y, p_z)/p_s$, and $\alpha = \omega/\omega_{\mathrm{c}}$ have
been introduced. The expression of the Coulomb potential, third line
in Eq. (\ref{eq:Homega-dl}), allows to obtain more compact formulae 
later in the present section. 

Next, given the equilibrium configuration
$(\vec{r}_0,\vec{0})$ of each ion, 
we expand the Hamiltonian (\ref{eq:Homega-dl}) 
to second order in the spatial displacement $\mathbf{q} = \mathbf{r} -
\mathbf{r}_0$ and $\mathbf{p}$ around zero, namely

\begin{eqnarray}
H_R(\vek{p},\vek{q}) \simeq H_R(\vek{0},\vek{r}_0) 
+ \frac{1}{2}\mathbf{d} \tilde{H}_R\mathbf{d}^{\mathsf{T}},
\label{eq:quadExp} 
\end{eqnarray}
where $\mathbf{d}^{\mathsf{T}}$ is the transpose of the row vector 
$\mathbf{d} \equiv
(q_{1,x},p_{1,x},q_{1,y},p_{1,y},\dots,q_{N,z},p_{N,z})$, 
and $\tilde{H}_R=\tilde{H}_R(\vek{0},\vek{r}_0)$ is the
Hessian matrix. Its non-zero matrix elements are given by:

\begin{eqnarray}
\frac{\partial^2 H_R}{\partial p_{\eta_k}^2} = 1,\qquad
\frac{\partial^2 H_R}{\partial p_{x_k}\partial y_k} = - 
\frac{\partial^2 H_R}{\partial p_{y_k}\partial x_k} = \alpha + \frac{1}{2},
\nonumber
\end{eqnarray}
\begin{eqnarray}
\frac{\partial^2 H_R}{\partial \eta_k\partial \mu_j} &=& 
\left[
1 - 2 \alpha_z^2 + (6 \alpha_z^2 - 1) \delta_{\eta,z}
\right]\frac{\delta_{\eta,\mu}\delta_{k,j}}{4}\nonumber\\
&+&\lim_{\epsilon \rightarrow 0}\sum_{s=1}^N
\frac{(1 - \delta_{k,s})[\delta_{s,j} + (1 -
\delta_{s,j})\delta_{|k-j|,0}]}
{\vert \vec{r}_k - \vec{r}_s + \epsilon\vert^3}\times\nonumber\\
&\times&(-1)^{\delta_{k,j}}\left[
\delta_{\eta,\mu}
-3
\frac{(\eta_k - \eta_s)(\mu_k - \mu_s)}{\vert\vec{r}_k - \vec{r}_s + \epsilon
  \vert^2}
\right],
\nonumber
\end{eqnarray}
where $\eta,\mu = x,y,z$, and $k,j=1,\dots,N$. 

\subsection{Symplectic diagonalization and canonical quantization}
\label{sec:symplettic}

Hereafter we utilize the frame $F_3$ that rotates at the frequency 
$\omega_{\mathrm{r}}$. Hence, we are allowed to 
drop $H_R(\vek{0},\vek{r}_0)$ in Eq. (\ref{eq:quadExp}) and the full
Hamiltonian reduces to the $6N\times6N$-matrix $H_R(\omega_{\mathrm{r}}) 
= \mathbf{d} \tilde{H}_R \mathbf{d}^{\mathsf{T}}/2$.

In order to perform the canonical quantization, we have first to transform 
the classical Hamiltonian $H_R(\omega_{\mathrm{r}})$ in canonical form. 
A transformation $S: (\vek p,\vek q)\rightarrow(\vek P, \vek Q)$ 
is canonical when the condition $S \mathbb J S^T = \mathbb J$ is 
satisfied, where $\mathbb J = i\bigoplus_{i=1}^{3N} \hat\sigma^y$ \cite{Fasano2006}. 
Since the Hessian matrix $\tilde{H}_R$ is real and positive definite,
Williamson's theorem~\cite{Williamson1936} guarantees that 

\beq
S \tilde{H}_R S^{\mathsf{T}} = W = 
\left(
\begin{array}{ccccc}
\omega_1 & & & & \\
 & \omega_1 & & & \\
 &  & \dots & & \\
 &  &  & \omega_{3N} & \\
 &  &  &  & \omega_{3N}
\end{array}
\right),
\eeq
where $\omega_k$ are real and positive numbers $\forall k=1,\dots,3N$,  
and $W$ is called the ``Williamson form'' of $\tilde{H}_R$. 

Given that, we can recast the classical Hamiltonian as

\beq
H_R(\omega_{\mathrm{r}}) = \frac{1}{2}\sum_{k=1}^{3N}\omega_k \Lambda_{2k-1}^2 
+ \frac{1}{2}\sum_{k=1}^{3N}\omega_k \Lambda_{2k}^2,
\eeq
where the new coordinates are determinated by the transformation
$\boldsymbol{\Lambda}^{\mathsf{T}} = (S^{-1})^{\mathsf{T}}
\mathbf{d}^{\mathsf{T}}$. For the sake of simplicity, hereafter, we use the definitions 
$Q_k := \Lambda_{2k-1}$ and $P_k := \Lambda_{2k}$ $\forall k=1,\dots,3N$. 
Thus, the Hamiltonian reduces to 

\beq
H_R(\omega_{\mathrm{r}}) = \frac{1}{2}\sum_{k=1}^{3N}\omega_k (Q^2_k + P^2_k),
\eeq 
that is, a sum of uncoupled harmonic oscillators. 

Similarly to Sec. \ref{sec:phonons}, we perform the canonical
quantization by promoting $Q_k, P_k$ to operators 
such that $[\hat Q_k,\hat P_s] = i\delta_{k,s}$. Besides this, we 
introduce the operators $\hat a_k = (\hat Q_k + i \hat P_k)/\sqrt{2}$, 
$\hat a_k^\dag = (\hat Q_k - i \hat P_k)/\sqrt{2}$ with $[\hat a_k,\hat a_s^\dag] = \delta_{k,s}$. 
Hence, the quantized Hamiltonian is simply given by

\beq
\hat H_R(\omega_{\mathrm{r}}) = \sum_{k=1}^{3N}\omega_k\left(\hat a_k^\dag \hat a_k 
+ \frac{1}{2}\right),
\eeq 
and we note that the eigenvalues $\omega_k$ are dimensionless. 

Finally, we rewrite the coupling between the ions and the
inhomogeneous laser field. The displacement of the ion 
from its equilibrium position can be written as 

\beq
\hat d_{j} = \frac{1}{\sqrt{2}}\sum_{k=1}^{3N} A_{k,j}^* \hat a_k + A_{k,j} \hat a_k^\dag
\eeq
with $A_{kj}=S_{2k-1,j} + i S_{2k,j}$, and where $j$ is an odd integer [see
the definition of the vector $\mathbf{d}$ after Eq. (\ref{eq:quadExp})]. 
Then, the matter-field interaction has the following expression

\begin{eqnarray}
\hat V \!& = &\! \sum_{j=1}^N [\vec{q}_j \cdot \vec{f}_j(t)] \hat\sigma^z_j 
       \!= \! \sum_{j=0}^{N-1} \hat\sigma^z_{j+1} \sum_{n=1}^3 \mathcal{F}_{3j+n}(t) \hat d_{2(n+3j)-1} \nonumber\\
       & = & \!\sum_{k=1}^{3N} \alpha_k^* \hat a_k + \alpha_k \hat a_k^\dag, 
\end{eqnarray}
where $\boldsymbol{\mathcal{F}}=(f_{1,x},f_{1,y},f_{1,z},\dots,f_{N,x},f_{N,y},f_{N,z})$, and 

\beq
\alpha_k = \frac{1}{\sqrt{2}}\sum_{j=0}^{N-1} \hat\sigma^z_{j+1} \sum_{n=1}^3 \mathcal{F}_{3j+n}(t)A_{k,2(n+3j)-1}.
\label{eq:alpha}
\eeq
Thus the full Hamiltonian is: $\hat H = \hat H_R(\omega_{\mathrm{r}}) + \hat V = \sum_k \hat H_k$, where 
$\hat H_k = \omega_k\left(\hat a_k^\dag \hat a_k + \frac{1}{2}\right) + \alpha_k^* \hat a_k + \alpha_k \hat a_k^\dag$.

\subsection{Two-qubit phase gate}
\label{sec:phasegate}

The time evolution of a phonon mode state, governed by the 
Hamiltonian $\hat H_k$, and a generic two-qubit state is 

\beq
\ket{\Psi_k;\Phi_{\mathrm{qbit}}(t)} = 
e^{-i\phi_k(t)}\hat{\mathfrak{D}}[\beta_k(t)]e^{-i\hat H_k^0}\ket{\Psi_k;\Phi_{\mathrm{qbit}} (0)},
\label{eq:Ut}
\eeq
where $\hat{\mathfrak{D}}[\beta_k(t)]$ 
is the displacement operator \cite{Gardiner2004}, $\beta_k(t)  =  -i \int_0^t\mathrm{d}s \alpha_k(s) e^{i\omega_k(s-t)}$, 
and $\hat H_k^0  =  \omega_k\left(\hat a_k^\dag \hat a_k + \frac{1}{2}\right)$. 

In order to disentangle the external dynamics due to the phonons 
and the internal  dynamics of the qubit states at the end of the 
gate operation, $t=\tau_{\mathrm{g}}$, the following condition 
has to be satisfied \cite{Garcia2005}

\begin{eqnarray}
\mathcal{I}_k = \frac{1}{\sqrt{\omega_k}}
\int_0^{\tau_{\mathrm{g}}}\mathrm{d} t \,e^{i\omega_k t}\alpha_k(t) = 0 
\qquad \forall k.
\label{eq:condforce}
\end{eqnarray}
This condition, however, is more general than the adiabatic elimination 
we performed in Sec. \ref{sec:adiafastgate}, whose aim was to highlight the difference in the 
accumulated two-particle phases among the most common quantum gate schemes based on 
pushing forces with off-resonant lasers. 

The necessary lateral force on the $j$-th and $k$-th ion,
$\vert\vec{f}_j \vert= \vert\vec{f}_k \vert= \mathcal{A}_P\hbar\omega_{xy}\cos(\nu t)
e^{-t^2/\tau^2_{\mathrm{g}}}/\vert \vec{r}_j^{\,0}-\vec{r}_k^{\,0}\vert$,  
is determined by setting the dimensionless parameter $\mathcal{A}_P$ to 
achieve a $\pi$ phase between the chosen pair of qubits. 
Then the fidelity is given by

\begin{widetext}
\begin{eqnarray}
\!\!\!\!\!F = \min_{\Phi^{\prime}_{\mathrm{qbit}}}\left\{
\mathrm{Tr}_{\mathrm{ph}}\left[\bra{\Phi^{\prime}_{\mathrm{qbit}}}
\hat{\mathfrak{U}}(t)\left(\hat\rho_T(0)\otimes\ket{\Phi_{\mathrm{qbit}}}
\bra{\Phi_{\mathrm{qbit}}}\right)\hat{\mathfrak{U}}^{\dagger}(t)\ket{\Phi^{\prime}_{\mathrm{qbit}}}\right]
\right\} = 
\min_{\pm}\prod_{k}\exp\left[
-\frac{\mathcal{A}_P^2}{4}\left(\frac{\vert\mathcal{I}_k^{(j_1)}\pm\mathcal{I}_k^{(j_2)}\vert^2}{1-e^{-\hbar \omega_k/k_B T}}\right)
\right],
\label{eq:fidelity}
\end{eqnarray}
\end{widetext}
where $\hat{\mathfrak{U}}(t)$ is the unitary evolution operator defined 
through Eq. (\ref{eq:Ut}), $\hat\rho_T(0)$ is the initial (canonical) density operator 
of the phonon modes at temperature $T$, $\ket{\Phi_{\mathrm{qbit}}}$ 
is the initial two-qubit state, and 

\begin{eqnarray}
\ket{\Phi^{\prime}_{\mathrm{qbit}}} = \sum_{\epsilon_1,\epsilon_2=0}^1(-1)^{\epsilon_1\epsilon_2}
c_{\epsilon_1,\epsilon_2}\ket{\epsilon_1}\ket{\epsilon_2}
\end{eqnarray}
is the desired logical target state we aim to attain. The integral 
$\mathcal{I}_k^{(j_{q})}$ for $q=1,2$ is given in Eq. (\ref{eq:condforce}) 
where the apex $(j_{q})$ refers to the ion we are considering, that is, 
$j=j_q$ in the sum of Eq. (\ref{eq:alpha}). 

Since we aim to achieve $\tau_{\mathrm{g}}\ll2\pi/\omega_{\mathrm{r}}$, 
we outline the following program: Firstly, we analyze 
the dependence of $\omega_{\mathrm{r}}$ on the total angular 
momentum $P_{\theta}$. This is achieved by fixing a priori a value 
of $P_{\theta}$ and then by determining the equilibrium configuration 
of the crystal, namely the positions and momenta of each ion (the 
most difficult part of the program). Since the crystal is a 
rigid body, it holds $p_{\theta_k}=m r_k^2\dot \theta_k + e r_k A_{\theta}(r_k) 
= m r_k^2(\omega_{\mathrm{r}} + \omega_{\mathrm{c}}/2)$ \cite{Dubin1999}, 
and from this relation the rotation frequency is extracted. Such an analysis allows us to 
find the smallest value of $\omega_{\mathrm{r}}$ such that 
$\beta<\beta_{\mathrm{c}} = 0.665/\sqrt{N}$ is fulfilled, that is, 
a 2D Wigner crystal configuration \cite{Dubin1999}. Then we choose the 
value of both $\omega_{\mathrm{c}}$ and $\nu$ in order to 
achieve high gate fidelity for a wide range of temperatures. 

The determination of the classical ground state is a 
multidimensional minimization constrained problem for which 
no deterministic and efficient algorithm is known. Here we used 
a variant of the Metropolis \cite{Metropolis1953} and the multidimensional 
constrained Newton algorithm like the one of Ref. \cite{Schweigert1995}. 
The first method allows us to sample randomly the relevant phase space 
region by choosing a slow decay of the acceptance probability and 
by using several annealing cycles. We then coarse-grained the obtained 
annealing trajectories into intervals, and we employed, for the lowest energy 
configuration on each interval, a Newton algorithm, which is very efficient 
in finding a local minimum provided that the initial value is already very close.
We have checked the reliability 
of our numerical energy minimization for $P_{\theta}=0$, that is 
$\omega_{\mathrm{r}} = 2 \omega_{\mathrm{c}}$, by comparing the results 
of Ref. \cite{Schweigert1995} for the minimal excitation frequency for several 
numbers $N$ of ions. 

\begin{figure}[t]
\begin{center}
\includegraphics{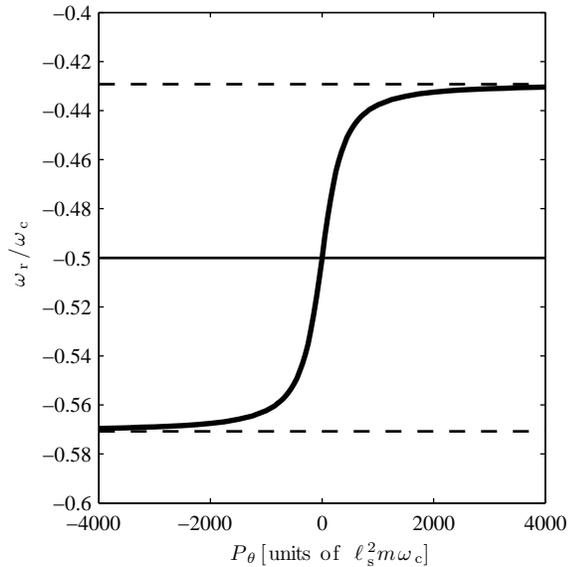}
\end{center}
\caption{(Color online). Ratio $\omega_{\mathrm{r}}/\omega_{\mathrm{c}}$ vs. 
total angular momentum $P_{\theta}$ for $\alpha_z=0.7$ and $N=30$. 
For $P_{\theta}=0$ we retrieve the well-known limit  
$\omega_{\mathrm{r}} = 2 \omega_{\mathrm{c}}$, in which there is 
no magnetic field in the rotating frame.}
\label{fig:wrL}
\end{figure}

\begin{figure}[t]
\begin{center}
\includegraphics{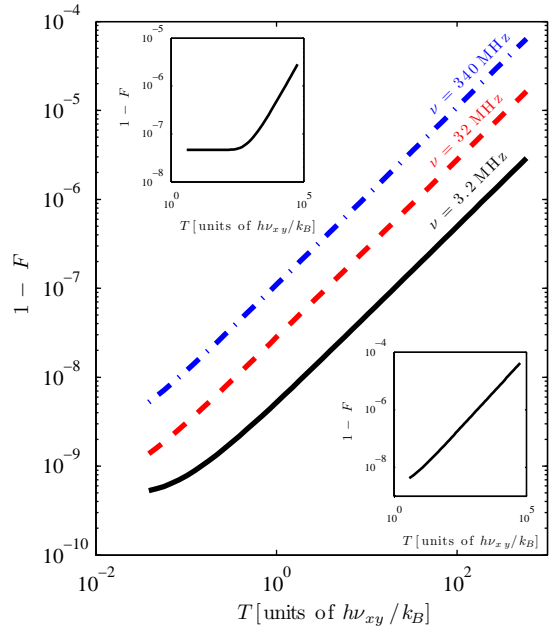}
\end{center}
\caption{(Color online). Infidelity vs. temperature for $\alpha_z=0.7$, $N=30$, 
and $P_{\theta} = 4\times 10^{3} \,\ell^2_{\mathrm{s}} m \omega_{\mathrm{c}}$. 
Parameters: $\nu_{\mathrm{c}} = \omega_{\mathrm{c}} / (2 \pi) =  76.08$ kHz, 
$\nu_{xy} = \omega_{xy}/(2\pi) = 5.38$ kHz, $\nu_z = \omega_z/(2\pi) = 53.26$ kHz, and 
$\nu_{\mathrm{r}} = \omega_{\mathrm{r}}/(2\pi) = 32.75$ kHz. The black (solid) line corresponds to 
$\tau_{\mathrm{g}}/\tau_{\mathrm{r}}=10^{-1}$ ($\tau_{\mathrm{g}} = 3 \,\mu$s), the red (dashed) line to 
$\tau_{\mathrm{g}}/\tau_{\mathrm{r}}=10^{-2}$ ($\tau_{\mathrm{g}} = 0.3 \,\mu$s), and the blue (dashdot) line to 
$\tau_{\mathrm{g}}/\tau_{\mathrm{r}}=10^{-3}$ ($\tau_{\mathrm{g}} = 0.03 \,\mu$s), with 
$\tau_{\mathrm{r}}=2\pi/\omega_{\mathrm{r}}$. The inset (on the left corner - top) provides the 
infidelity for $\tau_{\mathrm{g}}/\tau_{\mathrm{r}}=10^{-1}$ ($\tau_{\mathrm{g}} = 0.03 \,\mu$s) with 
$\nu_{\mathrm{c}} =  7.61$ MHz as in Ref. \cite{Mitchell2001}, 
$\nu= 300$ MHz, $\nu_{xy} = 537.97$ kHz, 
$\nu_z = 5.33$ MHz, and $\nu_{\mathrm{r}} = 3.27$ MHz. 
The inset (on the right corner - bottom) illustrates, for the same trapping 
parameters as for the former inset but with $\nu = 2.4$ MHz, the infidelity 
for $\tau_{\mathrm{g}}/\tau_{\mathrm{r}}=10$ ($\tau_{\mathrm{g}} = 3 \,\mu$s). 
Such modulation frequency $\nu$ lies within the gap among radial and 
$\vec{E}\times\vec{B}$ phonon modes (see Fig. \ref{fig:modes}).}
\label{fig:infid}
\end{figure}

We investigated the robustness of the modulated-carrier phase 
gate against temperature for a moderate number of ions $N=30$ and 
$\alpha_z = 0.70$. 
In Fig. \ref{fig:wrL} the dependence of the crystal rotation frequency 
on the total canonical angular momentum is showed, whereas in Fig. \ref{fig:infid} 
the gate infidelity for different values of the ratio $\tau_{\mathrm{g}}/\tau_{\mathrm{r}}$ 
is displayed. The results of Fig. \ref{fig:infid} refer to 
$P_{\theta} = 4000 \,\ell^2_{\mathrm{s}} m \omega_{\mathrm{c}}$, 
for which we obtain the smallest value of $\vert\omega_{\mathrm{r}}\vert$ 
in Fig. \ref{fig:wrL}. Beside this, we have for such a choice 
$\beta = 3.4\times 10^{-4}$, whereas $\beta_{\mathrm{c}} = 0.12$, 
that is, a stable 2D hexagonal lattice configuration. Given that, 
Fig.~\ref{fig:infid} shows that in order to reduce by a factor 10 
the ratio $\tau_{\mathrm{g}}/\tau_{\mathrm{r}}$ the fast modulation 
frequency $\nu$ of the force has to be (roughly) enhanced by a factor 10 as 
well. We also remark, that the three lines in Fig.~\ref{fig:infid} show 
an infidelity that is smaller for large gate operation times. The goal of the 
plot is to show how the modulation frequency increases when the ratio 
$\tau_{\mathrm{g}}/\tau_{\mathrm{r}}$ is reduced for an infidelity smaller 
than $10^{-4}$. Of course, by carefully tuning $\nu$ 
one can easily get a smaller infidelity for faster gates.

In the inset (left corner - top) the result of the gate infidelity for a cyclotron frequency 
100 times higher is showed, that is, the same 7.608 MHz of the 
experiment of Ref. \cite{Mitchell2001}. Here there are two important 
features to be highlighted: firstly, the gate fidelity is more robust for 
a wide range of temperatures with respect to the previous case 
where $\omega_{\mathrm{c}}/(2\pi) = 76.08$ kHz has been considered. 
On the other hand, already for $\tau_{\mathrm{g}}/\tau_{\mathrm{r}} = 0.1$, 
the frequency $\nu$ is on the order of hundreds of MHz. 
In the inset on the right (bottom) we show the gate infidelity again for the 
$\omega_{\mathrm{c}}=$7.608 MHz but for a smaller modulation frequency 
$\nu = 2.4$ MHz that lies in the gap between the two bands of different radial modes 
(the so called $\vec{E}\times\vec{B}$ and cyclotron modes. See also Fig. \ref{fig:modes}). 
In this scenario $\tau_{\mathrm{g}}/\tau_{\mathrm{r}} = 10$ and therefore a co-rotating 
laser beam is required. In conclusion we see that for 
$\omega_z\sim \omega_{\mathrm{c}}/\sqrt{2}$ if we desire to avoid the 
employment of a co-rotating laser beam the only possible way is to achieve very high 
frequencies for the modulation of the state dependent force. 

\begin{figure}[t]
\begin{center}
\includegraphics{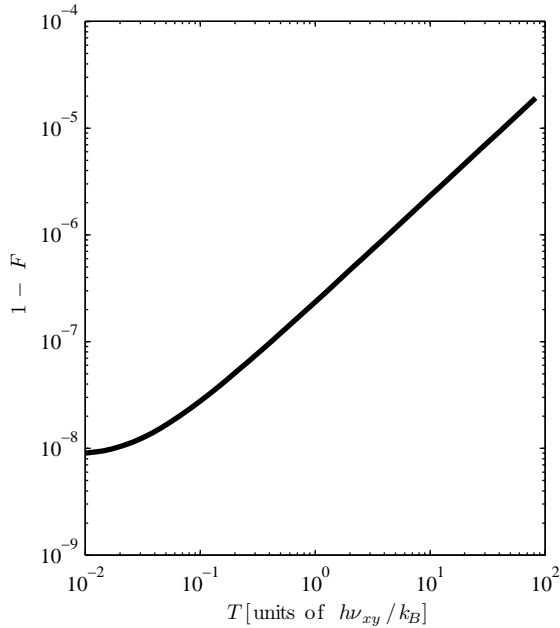}
\end{center}
\caption{(Color online). Infidelity vs. temperature for $\alpha_z=0.02$, $N=30$, 
and $P_{\theta} = 1.3\times 10^{5} \,\ell^2_{\mathrm{s}} m \omega_{\mathrm{c}}$. 
Parameters: $\nu_{\mathrm{c}} =  7.61$ MHz, 
$\nu_{xy} =  3.80$ MHz, $\nu_z =  152.16$ kHz, and 
$\nu_{\mathrm{r}} = 1.65$ kHz (see text for more details).}
\label{fig:infid_a02}
\end{figure}

\begin{figure*}[htb]
\centerline{
\includegraphics[width=120mm]{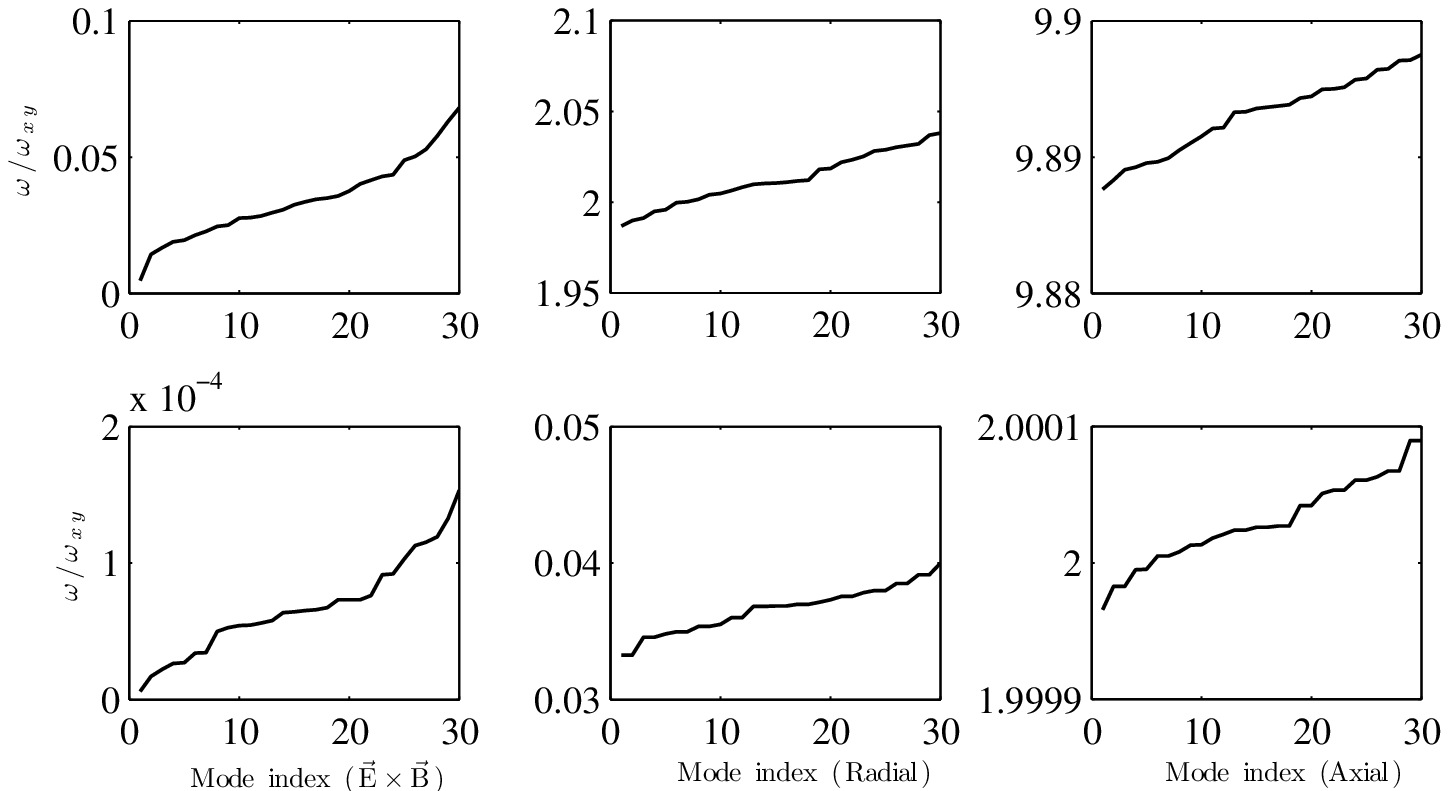}
}
\caption[]{(Color online). Modes for the parameters considered in Fig. \ref{fig:infid} 
(top row, $\alpha_z=0.70$, $P_{\theta} = 4\times 10^{3} \,\ell^2_{\mathrm{s}} m \omega_{\mathrm{c}}$) 
and Fig. \ref{fig:infid_a02} 
(bottom row, $\alpha_z=0.02$, $P_{\theta} = 1.3\times 10^{5} \,\ell^2_{\mathrm{s}} m \omega_{\mathrm{c}}$).
}
\label{fig:modes}
\end{figure*}

Alternatively, one can consider a smaller value of $\alpha_z$, which basically shifts 
upwards the graph of Fig. \ref{fig:wrL}, that is, by displacing the minimum of the 
$\vert\omega_{\mathrm{r}}\vert$ closer to zero for large values of $P_{\theta}$. 
This is the situation depicted in Fig. \ref{fig:infid_a02} for $N=30$, and $\alpha_z = 0.02$.  
Here it is possible to achieve gate operation times on the order of few $\mu$s 
with significantly smaller values of the modulation frequency. In the figure $\nu$ 
lies in the gap among axial and radial modes (see Fig. \ref{fig:modes}). Furthermore 
with $\tau_{\mathrm{g}}/\tau_{\mathrm{r}} = 6\times 10^{-3}$ we do not need a 
co-rotating laser beam.  This result is quite interesting since it works in a range 
of parameters that are currently employed in experiments (e.g., \cite{Biercuk2009}). 
Finally we also note that in this scenario $\omega_{xy} \gg \omega_z$, 
which is opposite to the requirement we identified in the case of larger $\alpha_z$ when 
$2\omega_r = \omega_c$. We note, however, that $\omega_{xy}$ is not the actual 
radial frequency when $\omega_{\mathrm{c}}\ne2\omega_{\mathrm{r}}$. 
Indeed, as shown in Eq. (\ref{eq:Homega}), the centrifugal potential (i.e., $-m\omega^2r^2/2$) modifies the confinement. 
Let us write $\omega_{\mathrm{r}}= (\omega_{\mathrm{c}} - \delta\omega)/2$, where $\delta\omega>0$. 
Then by substituting such definition into the second line of (\ref{eq:Homega}) we obtain an effective 
radial frequency given by: $\omega_{xy}^{\mathrm{eff}} = \frac{1}{2}\sqrt{\omega_{\mathrm{c}}^2 - \delta\omega^2 -2\omega_z^2}$. 
With the parameters of Fig. \ref{fig:infid_a02} we get $\omega_{xy}^{\mathrm{eff}}/(2\pi) = 31.47$ kHz 
which is significantly smaller than $\nu_z$ ($\sim 152$ kHz), and therefore the 2D lattice configuration 
is ensured. This fact is also confirmed by $\beta = 4\times 10^{-2}<\beta_{\mathrm{c}}$. 

Now, let us examine what is the required laser power in order to 
realize the gate and investigate the influence of scattered photons on the 
gate performance. 
A pair of narrow-waist ($\le 2\,\mu$m) adjacent laser beams in the 
standing-wave configuration produces the necessary force to be 
applied to each ion. Beside this, because of the tight focusing it reduces 
spontaneous emission and laser power.  Following the treatment 
of Ref. \cite{Sasura2003}, an estimate of the needed laser power to 
realize the logical gate is given by

\beq
P = \mathcal{A}_P\frac{\omega_{xy}\Delta \hbar c \kappa^2
w^2\sin^2(\gamma/2)}{3\Gamma\vert\vec{r}_i^{\,0} - \vec{r}_j^{\,0}\vert}.
\eeq
Here $\Delta=\omega_L - \omega_A$ is the detuning, that is, the 
difference between the laser  and the relevant atomic transition frequencies, 
$\Gamma$ is the linewidth of the transition, $\kappa = 2\pi/\lambda_L$ 
is the wave number with the laser wavelength $\lambda_L=2\pi c/\omega_L$, 
$c$ is the speed of light, $w$ is the size of the beam waist, and 
$\gamma$ is the angle between the $\kappa$ vectors of the two 
laser beams (see also Fig. \ref{fig:artistsview}). Additionally, we can estimate the influence of photon 
scattering on the gate fidelity as: $F_{\mathrm{scat}}=e^{-N_{\mathrm{phot}}}$, 
where the number of scattered photons in the standing-wave configuration 
is given by

\beq
N_{\mathrm{phot}} \approx \frac{\sqrt{2}\pi^3\epsilon_0 c\,m^2 w^2
 \omega_{xy}^4\vert\vec{r}_i^{\,0} - \vec{r}_j^{\,0}\vert^3}
{3e^2\lambda_L P}\sin\left(\frac{\gamma}{2}\right).
\eeq
As the last two formulae show, by adjusting $\gamma$ we can reduce 
the required laser power, but at the expenses of a larger number of 
scattered photons, and therefore of a worsening of the gate performance. 

\section{Modulated and state-dependent dipole force}
\label{sec:force}

In order to realize our quantum phase gate, $\ket{\epsilon_1,\epsilon_2}\rightarrow e^{i\theta\epsilon_1\epsilon_2}
\ket{\epsilon_1,\epsilon_2}$ with $\epsilon_{1,2} = 0,1$, we have to engineer the 
$\theta_{kj}$ phases in $\theta$ (see also Sec. ~\ref{sec:modcargate}). It is natural to 
demand that the desired value of $\theta$ is obtained with the smallest 
possible value of the applied force (i.e., laser power) or, alternatively, in 
the shortest possible time. This is equivalent to maximize 
$\theta$ by maximizing the effect of each $\theta_{kj}$. This happens when the phases $\theta_{01}$ and 
$\theta_{10}$ have the opposite sign with respect to the phases $\theta_{00}$ 
and $\theta_{11}$. Such condition is met when the applied force to the $j$-th 
ion satisfies the relation 

\beq
\vec{f}_j^{\,\ket{0}} = -\vec{f}_j^{\,\ket{1}}.
\label{eq:oppforces}
\eeq
 
Additionally, a necessary condition for the implementation of a 
modulated-carrier quantum phase gate is that the mean force acting on each
ion (respectively each of the modes) has to be zero over $\tau_{\mathrm{g}}$, 
that is, we have to fulfill Eq. (\ref{eq:condforce}). Such a requirement can be 
accomplished by making the modulation time symmetric around the center 
of the envelope of the laser pulse. To this aim, we impose the further condition 
on the force:

\beq
\int_0^\tau\mathrm{d}t \,\vec{f}_j^{\,\ket k}(t) = 0 \qquad \forall k=0,1,
\label{eq:meanzeroforce}
\eeq
where $\tau=2\pi/\nu$ is one period of the modulation. With such a condition we 
obtain a (fast) sinusoidal modulation of the force. Experimentally, 
this can be achieved, for example, with an acousto-optical 
modulator, which can vary the frequency of the laser light very quickly.

\subsection{Energy shifts}

\begin{figure}[t]
\begin{center}
\includegraphics[width=3.3in]{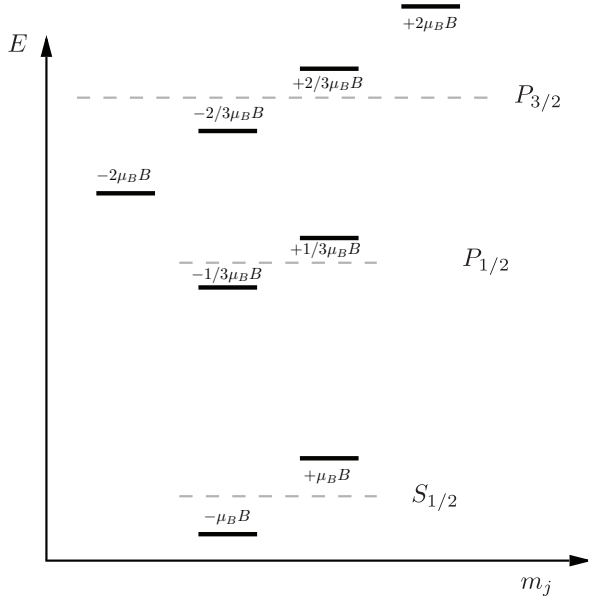}
\end{center}
\caption{(Color online). Energy shifts due to the Zeeman effect vs. the $m_j$ quantum number. 
The distances among the energy levels are not in scale.}
\label{fig:ZeemanShifts}
\end{figure}

In table \ref{tab:ZPBregimes} we provide for some ion 
species the energy splitting between the $P_{1/2}$ and $P_{3/2}$ levels in the absence of an 
external magnetic field together with the maximal value of magnetic field $B_Z$, under which the 
(normal) Zeeman limit can be applied, and the minimal value $B_{PB}$ above which we enter in 
the Paschen-Back regime. As we can gather from the table, the higher the atomic number of the ion
(or neutral atom) is, the larger $\Delta E$ and the limits $B_Z$, $B_{PB}$. For instance, for the infidelity results we 
showed in the previous section, the corresponding magnetic field at $\omega_{\mathrm{c}} = 7.608$ 
MHz are: $B=4.5$ T and $B=12$ T for Beryllium and Magnesium, respectively. These are also the 
values used in current experiments. Thus, for all alkaline-earth-metal atoms the Zeeman regime 
applies, and therefore $\hat H_B =\frac{\mu_{\mathrm{B}}}{\hbar} g_J\hat{J}_z B_z$ well describes 
the interaction of an ion with the external magnetic field. Here $g_J$ is the Land\'e factor \cite{Bethe1957} 
and the nuclear contribution has been neglected ($g_I\sim 10^{-3}$). Besides, since the external 
magnetic field has a strength of few Tesla, the ionic hyperfine structure can be also neglected.
Additionally, we note that in the Paschen-Back regime the transitions from the energy (split) ground state 
($S$-level) to the one of the excited levels ($P$) are identical for both ground levels when the ion is 
illuminated with a laser beam of a given polarization and frequency. Consequently, the dipole force 
(see Sec. \ref{sec:dipoleforce}) would be the same for both states, and therefore it would not be 
possible in such a regime to have state-dependent forces. Instead, this is not the case for the broken 
degeneracy of the $S$ and $P$ levels due to the Zeeman effect (see Fig. \ref{fig:ZeemanShifts}).

\begin{table}
\begin{tabular}{ccccccccccccccccccccccccccccccccccccc}
\hline
\hline
Atom/Ion & & & $\Delta E/\hbar$ (THz) & & & $B_Z$ (T) & & & $B_{PB}$ (T) \\
\hline
\hline
Be II & & & 1.239 & & & 7.043 & & & 28.170  \\
Mg II & & & 17.249 & & & 98.070 & & & 392.286  \\
Ca II & & & 41.985 & & & 238.711 & & & 954.845  \\
Na I & & & 3.242 & & & 18.410 & & & 73.651  \\
\hline
\hline
\end{tabular}
\caption{\label{tab:ZPBregimes} Energy splitting among the $P_{1/2}$ and $P_{3/2}$ levels without 
external the magnetic field (second column from the left). In the third and fourth columns (from the 
left) the maximal and minimal value of the external magnetic field, which fix respectively the upper 
and lower bound for the Zeeman and Paschen-Back regimes, are given. 
}
\end{table}

Finally, we note that given the selection rule $s - s^{\prime} = 0$ on the quantum number of the 
spin operator, optical transitions between the two levels of $S_{1/2}$ are not allowed. This fact 
reduces the possibility of undesired flips among the qubit pair, and therefore decoherence and 
dephasing mechanisms are strongly suppressed. Henceforth, we shall consider the lower energy 
level of $S_{1/2}$ as the logical state $\ket{0}$, whereas the upper one as $\ket{1}$.

\subsection{Dipole force and dipole matrix elements}
\label{sec:dipoleforce}

The dipole force is produced by an intensity gradient of the laser beam illuminating the atom, which is far detuned 
from the relevant atomic transition, whose levels are ac-Stark shifted. Such an energy shift creates an additional potential 
for the particle. For a two-level atom and in the large detuning limit $\Delta\gg\vert\Omega\vert$, the dipole force on 
the lower energy level reads

\beq
\vec{f} = -\frac{\hbar}{4\Delta}\nabla\vert\Omega(t,\vec{r})\vert^2,
\eeq
where the Rabi frequency $\Omega(t,\vec{r})$ on the atomic transition is given by

\begin{align}
\hbar \Omega &= - \vec{d}_{ab} \cdot \vec{\mathcal E}(\vec r, t) 
=- \vert\vec{d}_{ab}\vert \, \mathcal E_0 \, \chi(\vec r,t),
\end{align}
whereas the laser is assumed to be a classical light field. Here $\vec d_{ab}$ 
represents the matrix element of the dipole moment operator for the transition 
$\ket{a}\equiv\ket{j=1/2;m_j}\rightarrow\ket{b}\equiv\ket{j=1/2,3/2;m_j}$ for 
a given polarization of the electric field $\vec{\mathcal E}$ with strength 
$\mathcal E_0$, and $\chi(\vec r,t)$ is the spatial and temporal pulse shape. 

The bare (unshifted) detunings are then defined as: $\delta_{\mathrm{D}_1}= \omega_{L_1}-\omega_{\mathrm{D}_1}$ 
and $\quad \delta_{\mathrm{D}_2}= \omega_{L_2}-\omega_{\mathrm{D}_2}$, 
where $\omega_{L_1}$ and $\omega_{L_2}$ are the laser frequencies. 
In addition, in order to reduce the probability of unwanted photon scattering 
processes, we require that

\begin{align}
\abs{\mu_{\mathrm{B}} \,B} &\ll \abs{\delta_{\mathrm{D}_1}} \ll \Delta E & \abs{\mu_{\mathrm{B}} \,B} &\ll \abs{\delta_{\mathrm{D}_2}} \ll \Delta E.
\end{align}
In table \ref{tab:redblueforces} we provide the expressions of the state dependent forces for all
polarizations of the laser field. Since $j=1/2$ for all relevant transitions, hereafter, for the sake 
of simplicity, we shall denote the reduced matrix element by $\mathcal{M}_{\mathrm{D}_1} = \mathcal{M}_{1/2,1/2}$ 
and $\mathcal{M}_{\mathrm{D}_2} = \mathcal{M}_{1/2,3/2}$, where 
$\mathcal{M}_{jj^{\prime}}:=\langle j,m_j\parallel e \hat{r}\parallel j^{\prime},m_j^{\prime}\rangle$.

\begin{table*}
\begin{tabular}{cc*{4}{c}*{3}{c}}
\hline
\hline
 & & & & & & & & \vspace{-0.2cm}\\
$\mathrm{Polarization}$ & \hspace{0.3cm} & $\vec{f}^{\,\ket{0}}\, (\mathrm{D}_1)$ &
\hspace{0.3cm} & $\vec{f}^{\,\ket{1}} (\mathrm{D}_1)$ & \hspace{0.3cm} & $\vec{f}^{\,\ket{0}}\, (\mathrm{D}_2)$ 
& \hspace{0.3cm} & $\vec{f}^{\,\ket{1}}\, (\mathrm{D}_2)$ \vspace{0.2cm} \\
\hline
\hline
 & & & & & & & & \\
$\sigma^-$ & & 0 & & \large$-\frac{\mathcal{M}_{\mathrm{D}_1} \mathcal E_0^2 
\nabla \chi^2(\vec r,t)}{2\hbar (3 \delta_{\mathrm{D}_1}+4\mu_{\mathrm{B}} B/\hbar)}$   & & \large$-\frac{\mathcal{M}_{\mathrm{D}_2}\mathcal E_0^2 
\nabla \chi^2(\vec r,t)}{4\hbar(\delta_{\mathrm{D}_2}+\mu_{\mathrm{B}} B/\hbar)}$  & & \large$-\frac{\mathcal{M}_{\mathrm{D}_2} 
\mathcal E_0^2 \nabla \chi^2(\vec r,t)}{4\hbar(3\delta_{\mathrm{D}_2}+5\mu_{\mathrm{B}} B/\hbar)}$   \vspace{0.3cm} \\
$\pi$ & & \large$\frac{\mathcal{M}_{\mathrm{D}_1} \mathcal E_0^2 
\nabla \chi^2(\vec r,t)}{4\hbar(2\mu_{\mathrm{B}} B/\hbar - 3\delta_{\mathrm{D}_1})}$ & &  \large$-\frac{\mathcal{M}_{\mathrm{D}_1} \mathcal E_0^2 
\nabla \chi^2(\vec r,t)}{4\hbar(3\delta_{\mathrm{D}_1} + 2\mu_{\mathrm{B}} B/\hbar)}$  &  & \large$\frac{\mathcal{M}_{\mathrm{D}_2} \mathcal E_0^2 
\nabla \chi^2(\vec r,t)}{2\hbar(\mu_{\mathrm{B}} B/\hbar - 3\delta_{\mathrm{D}_2})}$ & & \large$-\frac{\mathcal{M}_{\mathrm{D}_2} \mathcal E_0^2 
\nabla \chi^2(\vec r,t)}{2\hbar(3 \delta_{\mathrm{D}_2}+\mu_{\mathrm{B}} B/\hbar)}$  \vspace{0.3cm} \\
$\sigma^+$ & & \large$\frac{\mathcal{M}_{\mathrm{D}_1} \mathcal E_0^2 
\nabla \chi^2(\vec r,t)}{2\hbar(4\mu_{\mathrm{B}} B/\hbar - 3 \delta_{\mathrm{D}_1})}$ &  & 0  &  & \large$\frac{\mathcal{M}_{\mathrm{D}_2} \mathcal E_0^2 \nabla 
\chi^2(\vec r,t)}{4\hbar(5\mu_{\mathrm{B}} B/\hbar - 3\delta_{\mathrm{D}_2})}$ & & \large$\frac{\mathcal{M}_{\mathrm{D}_2} 
\mathcal E_0^2 \nabla \chi^2(\vec r,t)}{4\hbar(\mu_{\mathrm{B}} B/\hbar - \delta_{\mathrm{D}_2})}$  \\
 & & & & & & & & \\
\hline
\hline
\end{tabular}
\caption{\label{tab:redblueforces} Dipole forces for all polarizations of the laser fields and internal (logical) states.
}
\end{table*}

\subsection{Different laser configurations}
\label{sec:laserconfigs}

\begin{figure}[t]
\begin{center}
\includegraphics[width=3.3in]{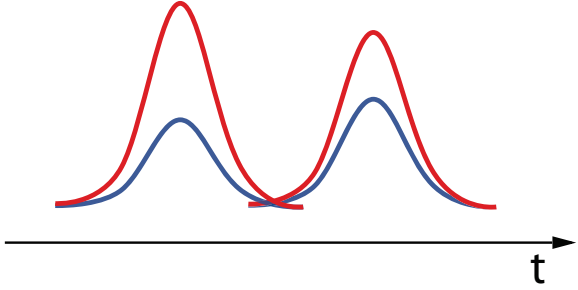}
\end{center}
\caption{(Color online). Sketch of the pulse sequence in order to design the necessary dipole forces to 
implement the modulated-carrier phase gate. The upper lines are red detuned, whereas the lower lines 
are blue detuned. In the figures the time is in arbitrary units.}
\label{fig:PulseSequence}
\end{figure}

As it is evident from the table \ref{tab:redblueforces}, using only a laser pulse either on the D$_1$ transition line or 
on the D$_2$ one it is not possible to fulfill the condition (\ref{eq:oppforces}). To this aim we need a second laser pulse 
(see Fig. \ref{fig:PulseSequence} on the left) with a different detuning from the first pulse. In this section we discuss 
several combinations of the laser polarization in order to satisfy both (\ref{eq:oppforces}) and (\ref{eq:meanzeroforce}). 

\subsubsection{Pulses with the same polarization}

\begin{figure}[t]
\begin{center}
\includegraphics[width=3.5in]{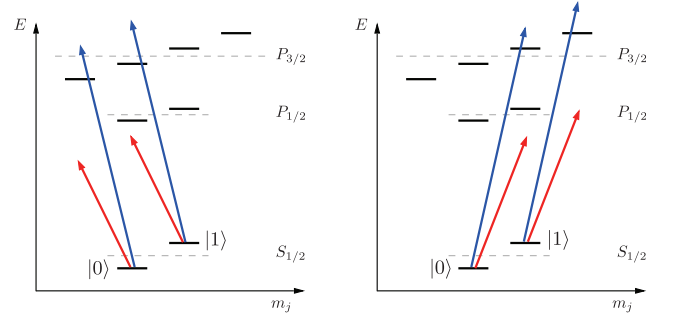}
\end{center}
\caption{(Color online). Modulated-carrier gate with the same polarization of the laser fields: 
$\sigma^-$-polarized (left), and $\sigma^+$-polarized (right). The distances among the energy 
levels are not in scale.}
\label{fig:PulseSeqEqualPol}
\end{figure}

In this case we first generate two laser pulses with different frequencies 
but with the same $\sigma^-$ polarization, as in Fig. \ref{fig:PulseSeqEqualPol}, which corresponds 
to the first sequence of pulses (on the left) in Fig. \ref{fig:PulseSequence}. Such a configuration of lasers 
yields the following state-dependent forces: 

\begin{align}
\vec f^{\,\ket 1} &= -\frac{\mathcal{M}_{\mathrm{D}_1} \mathcal E_{01}^2 
\nabla \chi^2(\vec r,t)}{2\hbar (3 \delta_{\mathrm{D}_1}+4\mu_{\mathrm{B}} B/\hbar)}
-\frac{\mathcal{M}_{\mathrm{D}_2} 
\mathcal E_{02}^2 \nabla \chi^2(\vec r,t)}{4\hbar(3\delta_{\mathrm{D}_2}+5\mu_{\mathrm{B}} B/\hbar)},
\nonumber\\
\vec f^{\,\ket 0} &= -\frac{\mathcal{M}_{\mathrm{D}_2}\mathcal E_{02}^2 
\nabla \chi^2(\vec r,t)}{4\hbar(\delta_{\mathrm{D}_2}+\mu_{\mathrm{B}} B/\hbar)}.
\end{align}
To simplify the notation we make the following replacements:  $\mathcal X_{\mathrm{D}_i} =  
\mathcal{M}_{\mathrm{D}_i}\mathcal E_{0i}^2 \nabla \chi^2(\vec r,t)$ and 
$\mathcal B = \mu_B B/\hbar$, where $\mathcal E_{0i}$ refers to the electric field 
strength of either the D$_1$ ($i=1$) or D$_2$ ($i=2$) line. Thus, in order to fulfill 
(\ref{eq:oppforces}), we have to solve the equation

\beq
\frac{\mathcal X_{\mathrm{D}_2}}{\delta_{\mathrm{D}_2}+\mathcal B} + \frac{ 
\mathcal X_{\mathrm{D}_2}}{3\delta_{\mathrm{D}_2}+5 \mathcal B}
+\frac{2\mathcal X_{\mathrm{D}_1}}{3\delta_{\mathrm{D}_1}+4 \mathcal B}=0 .
\eeq
This can be resolved for both the intensities or the detunings, so 
one of them can be considered as a given parameter. For the 
$\sigma^+$-polarization we obtain an analogue equation:

\beq
\frac{\mathcal X_{\mathrm{D}_2}}{\delta_{\mathrm{D}_2}-\mathcal B} 
+ \frac{\mathcal X_{\mathrm{D}_2}}{3\delta_{\mathrm{D}_2}-5 \mathcal B}
+\frac{2\mathcal X_{\mathrm{D}_1}}{3\delta_{\mathrm{D}_1}-4 \mathcal B}=0 .
\label{eq:example}
\eeq
As an example, we solve equation (\ref{eq:example}), for instance, for the intensities, 
and be obtain

\beq
\frac{\mathcal X_{\mathrm{D}_1}}{\mathcal X_{\mathrm{D}_2}} = 
\frac{(4 \mathcal B - 3 \delta_{\mathrm{D}_1})(2\delta_{\mathrm{D}_2} - 3 \mathcal B)}
{(\delta_{\mathrm{D}_2} - \mathcal B) (3 \delta_{\mathrm{D}_2} - 5 \mathcal B)}.
\label{eq:solexample}
\eeq
Such a solution, however, fulfills only the condition (\ref{eq:oppforces}) but not 
the one given by Eq. (\ref{eq:meanzeroforce}). To this aim we need an additional 
two-pulse sequence, as showed in Fig. \ref{fig:PulseSequence} on the right. 
Such two pulses can have different strengths of intensities and detunings, but 
they must have the same spatial and temporal profile $\chi({\vec r,t})$ of the 
first sequence. Again, we get, if we solve with respect to the intensities, a 
solution like the one given in Eq. (\ref{eq:solexample}), which in general 
will be different from the solution (\ref{eq:solexample}) for the first 
sequence of pulses. With such solution we can then easily satisfy 
also the mean zero force condition (\ref{eq:meanzeroforce}) by adjusting the 
ratios of either the intensities or the detunings. 

\begin{figure}[t]
\centering
\includegraphics{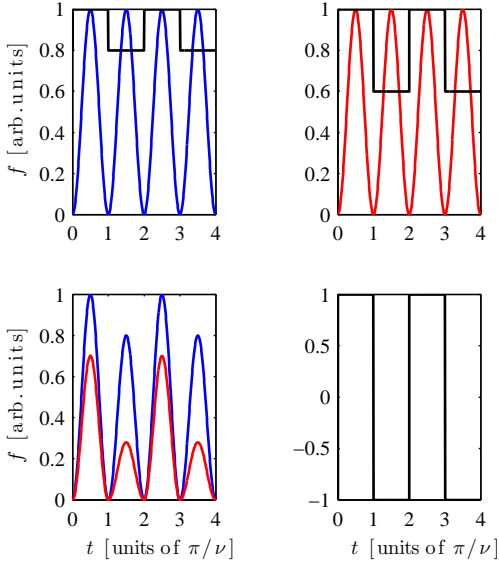}
\caption{\label{fig:pulses}(Color online). Designed laser modulation forces in order to fulfill both 
(\ref{eq:meanzeroforce}) and (\ref{eq:oppforces}). The blue detuned laser 
intensity with a superimposed square wave signal (top-left) and similarly for the red 
detuned one (top-right). The resulting signals are depicted on the bottom-left, whereas 
switching of polarization is given on the bottom-right.}
\end{figure}

In Fig. \ref{fig:pulses} we display a simple example that shows how to achieve the necessary 
laser pulse sequence. We modulate the intensities of the blue ($b$) and red ($r$) 
detuned laser signals like $I^{b (r)}(t) = I^{b (r)}_0 \sin^2(\nu t)$. The sequence starts 
($t=0$) with both lasers with $\sigma^+$-polarization and an intensity ratio 
$R_+ = I^{b}_0/I^{r}_0$ given by Eq. (\ref{eq:solexample}). Then, at time $t=\pi/\nu$, the 
polarization of the two laser fields is changed to $\sigma^-$ with another intensity 
ratio given by $R_- = I^{b}_0/I^{r}_0$. The ratio $R_-$ will differ from $R_+$, since 
in general the dipole moments are different for the two polarizations. Thus, 
by changing the polarization at each minimum of the laser intensity and by 
choosing the proper ratio $R_{\pm}$ we are able to fulfill the condition 
(\ref{eq:meanzeroforce}). In order to satisfy the condition (\ref{eq:oppforces}) 
we have to design furthermore the ratio of the two successive pulses. 
This is done by multiplying the two intensities $I^{b (r)}(t)$ with square wave 
signals which are displayed in Fig. \ref{fig:pulses} (top) by the black lines. The 
resulting pulses are showed in Fig. \ref{fig:pulses} (bottom-left) whose polarization 
state is depicted on the right lower corner. This procedure ensures that both the 
conditions (\ref{eq:meanzeroforce}) and (\ref{eq:oppforces}) are fulfilled. 

\subsubsection{Pulses with the different polarization}

\begin{figure}[t]
\begin{center}
\includegraphics[width=3.5in]{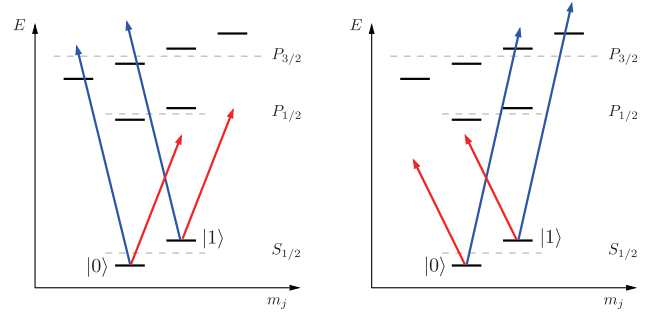}
\end{center}
\caption{(Color online). Modulated-carrier gate with different polarization of the laser fields. The distances among the energy 
levels are not in scale.}
\label{fig:PulseSeqDifferentPol}
\end{figure}

The situation in which the laser beams have different polarization is depicted in Fig. 
\ref{fig:PulseSeqDifferentPol}. If we illuminate the ion with red detuned and $\sigma_+$ 
polarized light and with a blued detuned and $\sigma_-$ beam (Fig. \ref{fig:PulseSeqDifferentPol} on the left)  
we have the following state-dependent forces

\begin{align}
\vec f^{\,\ket 1} &= -\frac{\mathcal{M}_{\mathrm{D}_2} 
\mathcal E_0^2 \nabla \chi^2(\vec r,t)}{4\hbar(3\delta_{\mathrm{D}_2}+5\mu_{\mathrm{B}} B/\hbar)},
\nonumber\\
\vec f^{\,\ket 0} &= \frac{\mathcal{M}_{\mathrm{D}_1} \mathcal E_0^2 
\nabla \chi^2(\vec r,t)}{2\hbar(4\mu_{\mathrm{B}} B/\hbar - 3 \delta_{\mathrm{D}_1})} 
-\frac{\mathcal{M}_{\mathrm{D}_2}\mathcal E_0^2 
\nabla \chi^2(\vec r,t)}{4\hbar(\delta_{\mathrm{D}_2}+\mu_{\mathrm{B}} B/\hbar)}, 
\end{align}
whereas for the inverted polarization sequence (Fig. \ref{fig:PulseSeqDifferentPol} on the right) 
we have

\begin{align}
\vec f^{\,\ket 1} &= \frac{\mathcal{M}_{\mathrm{D}_2} 
\mathcal E_0^2 \nabla \chi^2(\vec r,t)}{4\hbar(\mu_{\mathrm{B}} B/\hbar - \delta_{\mathrm{D}_2})} 
-\frac{\mathcal{M}_{\mathrm{D}_1} \mathcal E_0^2 
\nabla \chi^2(\vec r,t)}{2\hbar (3 \delta_{\mathrm{D}_1}+4\mu_{\mathrm{B}} B/\hbar)},
\nonumber\\
\vec f^{\,\ket 0} &= \frac{\mathcal{M}_{\mathrm{D}_2} \mathcal E_0^2 \nabla 
\chi^2(\vec r,t)}{4\hbar(5\mu_{\mathrm{B}} B/\hbar - 3\delta_{\mathrm{D}_2})}. 
\end{align}

Both the schemes with the same and with different polarization have the drawback 
that the transition to the excited level $P_{3/2}$ couples both the ground states of $S_{1/2}$, 
and therefore producing an additional force that has to be compensated with another laser 
beam. Apart from the technical difficulty of putting another laser beam, such a beam would 
also enhance the probability of promoting an ion to an excited level of $P_{3/2}$. Such 
excitation would cause an additional error during the course of the gate because of 
spontaneous emission. Indeed, the ion could decay either in the other qubit state or even 
worst, such as for the $D$-levels in calcium, in another metastable state, which would be 
useless for the purposes of QIP. 

\begin{figure}[t]
\begin{center}
\includegraphics[width=3.5in]{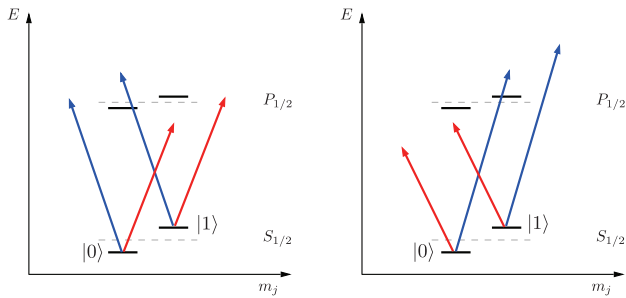}
\end{center}
\caption{(Color online). Variant of the modulated-carrier gate with different polarization of 
the laser fields. The distances among the energy levels are not in scale.}
\label{fig:PulseSeqDifferentPolBis}
\end{figure}

Given that, in order to avoid such scenario, we can make still use of the scheme illustrated 
in Fig. \ref{fig:PulseSeqDifferentPol}, but by avoiding the coupling to the $P_{3/2}$ manifold, 
as it is showed in Fig. \ref{fig:PulseSeqDifferentPolBis}. Here, however, we couple the ground 
state $S_{1/2}$ to only the manifold $P_{1/2}$. The pulse sequences are then the same as 
previously described for the other scheme. The detuning from the $P_{1/2}$ manifold, however, 
has to be carefully chosen, that is, it has to be much smaller than the energy difference among 
the $P_{1/2}$ and $P_{3/2}$ levels and much larger than $\mu_{\mathrm{B}} B$. Hence, such 
variant works well for sufficiently small magnetic fields. 

In principle there are other possible arrangements either by keeping the frequencies of the 
laser beams constant or by keeping constant their intensities. Such combinations rely also 
on the technical feasibility in an experimental setup. An important requirement for the design 
of such state-dependent forces is, however, the switch of the field polarization, which has to 
be very fast in order to fulfill the condition set by the Eq. (\ref{eq:meanzeroforce}), as it is 
also shown in the example of Fig. \ref{fig:pulses}(d). This can be experimentally accomplished 
with Pockels cells, which can be used to manipulate the polarization and the 
phase of the laser.




\section{Conclusions}
\label{sec:conc}

In this work we have analyzed in detail the implementation of the modulated-carrier 
gate presented for the first time in Ref. \cite{Taylor2008}. Firstly, we presented the 
underlying idea of the modulated-carrier gate and we provided details of the calculations 
that were only briefly mentioned in Ref. \cite{Taylor2008}. In that analysis the frame 
of reference rotates at the same frequency of the crystal rotation, whose frequency 
was set to $\omega_{\mathrm{c}} = 2 \omega_{\mathrm{r}}$. Within this setting the minimal 
coupling term in the many-body Hamiltonian vanishes. Such approach 
allows a straightforward canonical quantization of the many-body Hamiltonian, which 
reduces to a sum of $3N$ independent harmonic oscillators. Even though this situation 
greatly simplifies the numerical analysis it does not permit to fulfill the condition 
$\omega_{\rm r}\tau_{\mathrm{g}}/(2\pi) \ll 1$, which would avoid the utilization of 
a co-rotating laser and therefore simplifying the experimental realization of the 
proposed quantum hardware.  We thus have analyzed the situation 
in which $\omega_{\mathrm{c}} \ne 2 \omega_{\mathrm{r}}$. Within this scenario it 
is no longer possible to remove the minimal coupling term in the Hamiltonian 
of the Coulomb crystal. Nevertheless, by utilizing the Williamson theorem for positive 
definite matrices, we were able to diagonalize the classical many-body Hamiltonian, 
whose normal modes are a combination of both the position and momentum variables. 
As a consequence, we were able to perform the canonical quantization. The resulting 
(quantized) Hamiltonian is again given by a sum of independent harmonic oscillators. 
In this new situation, however, the matter-field interaction, responsible of the push on the ion, 
depends on both conjugate ``position" and ``momentum" operators. We proceeded further on by analyzing 
the performance of the quantum phase gate and we showed its robustness for a wide 
range of experimentally accessible temperatures. Importantly, we were able to demonstrate 
that such robustness is also displayed for a wide range of ratios $\tau_{\mathrm{g}}/\tau_{\mathrm{r}}$, 
therefore allowing to reduce up to three orders of magnitude the gate operation time 
compared to the previous analysis \cite{Taylor2008}. The drawback 
is that one has to enhance the modulation frequency $\nu$ up to 
hundreds of MHz in order to speed up the gate operation. We found, however, 
that by reducing the ratio $\omega_z/\omega_{\mathrm{c}}$, at large values of angular momentum 
it is possible to achieve small rotation frequencies such that $\omega_{\rm r}\tau_{\mathrm{g}}/(2\pi) \ll 1$ 
is fulfilled and high fidelity, for a broad range of temperatures, can be obtained 
with few MHz of modulation frequency. This result is quite promising since it has
been attained with a cyclotron frequency that is used in current experiments. 

Finally, we have provided a complete description for the design of the necessary forces to 
be applied on the ions in order to accomplish the desired quantum computation scheme. 
To this aim, we have analyzed the experimentally relevant region of external magnetic field. 
For all earth-alkali-metal ion species normally used in currents experiments the normal Zeeman effect 
provides, with good approximation, the right description of the energy shifts of the $S$ and $P$ 
levels. In addition, we have also analyzed several possible 
laser configurations and for each one we discussed advantages as well as drawbacks and, in 
some cases, we suggested alternative solutions. 

Further investigations of such a quantum computing proposal may rely on further optimization 
of both the force modulation together with a reduced gate operation time and its robustness against 
optimal pulse distortions \cite{Negretti2010}. This can be achieved 
by means of quantum optimal control techniques. Beside this, a detailed analysis, similar to Ref. 
\cite{Poulsen2010}, in order to characterize and quantify all types of errors coming from the 
quantum dynamics, especially due to nonlinearities in the ion-pushing force, will be pursued 
in future investigations.




\section*{Acknowledgments}

We are grateful to J. J. Bollinger for his critical reading of the manuscript. 
J.B. acknowledges G. De~Chiara and E. Kajari for helpful discussions, and 
A.N. useful correspondence with E. Pagani on symplectic transformations. 
We acknowledge financial support from the EU Integrated Project AQUTE, 
PICC (T.C.), the Deutsche Forschungsgemeinschaft within the Grant No. 
SFB/TRR21 (A.N.,T.C.), the Marie Curie Intra European Fellowship (Proposal Nr. 236073, OPTIQUOS) 
within the 7th European Community Framework Programme (A.N.), the 
Forschungsbonus of the University of Ulm and of the Ulmer Universit\"atsgesellschaft (A.N.), 
the Spanish Ministry of Science and Innovation (Consolider Ingenio
2010 ``QOIT'', QNLP FIS2007-66944), and the European Science Foundation
(EUROQUAM ``Cavity-Mediated Molecular Cooling") (J.B.).


\bibliography{IonPaperBib}
\end{document}